\documentclass[letterpaper,aps,prb,onecolumn,superscriptaddress]{revtex4}
\usepackage{setspace,subfigure}
\usepackage{amsmath,amssymb}
\usepackage{amsmath}
\usepackage{graphicx}
\usepackage[usenames,dvipsnames]{color}
\usepackage{colordvi}
\usepackage{float}

\begin{document}

\title{Polaron formation: Ehrenfest dynamics vs. exact results}
\date{\today }

\author{Guangqi Li}
 \affiliation{Department of Chemistry, Northwestern University, Evanston IL, 60208, USA}
 \author{Bijan  Movaghar}
 \affiliation{Department of Chemistry, Northwestern University, Evanston IL, 60208, USA}
 \author{Abraham Nitzan\footnote{nitzan@post.tau.ac.il}}
 \affiliation{School of Chemistry, Tel-Aviv University, Tel-Aviv 69978, Israel}
 \author{Mark A. Ratner\footnote{ratner@northwestern.edu}}
 \affiliation{Department of Chemistry, Northwestern University, Evanston IL, 60208, USA}
\begin{abstract}
\setstretch{1.5}
\large
We use a 1-dimensional tight binding model with an impurity site characterized by electron-vibration coupling, to describe electron transfer and localization at zero temperature, aiming to examine the process of polaron formation in this system. In particular we focus on comparing a semiclassical approach that describes nuclear motion in this many vibronic-states system on the Ehrenfest dynamics level to a numerically exact fully quantum calculation based on the Bonca-Trugman method [J. Bon\v{c}a and S.~A. Trugman, Phys. Rev. Lett. {\bf 75},  2566  (1995)]. In both approaches, thermal relaxation in the nuclear subspace is implemented in equivalent approximate ways: In the Ehrenfest calculation the uncoupled  (to the electronic subsystem) motion of the classical (harmonic) oscillator is simply damped as would be implied by coupling to a markovian zero temperature bath. In the quantum calculation, thermal relaxation is implemented by augmenting the Liouville equation for the oscillator density matrix with kinetic terms that account for the same relaxation. In both cases we calculate the probability to trap the electron in a polaron cage and the probability that it escapes to infinity. Comparing these calculations, we find that while both result in similar long time yields for these processes, the Ehrenfest-dynamics based calculation fails to account for the correct timescale for the polaron formation. This failure results, as usual, from the fact that at the early stage of polaron formation the classical nuclear dynamics takes place on an unphysical average potential surface that reflects the otherwise-distributed electronic population in the system, while the quantum calculation accounts fully for correlations between the electronic and vibrational subsystems. 
\end{abstract}

\maketitle

\large
\setstretch{1.5}
\section{Introduction}
Electron transfer (ET) between molecular systems has long been recognized as a key process in many research fields of chemistry, physics and biology\cite{marc85,barb96,bixo99,berlin04,skour10}. Many of its aspects are described by the Marcus theory\cite{marc56}, which has been extended to describe such areas as artificial solar-energy conversion\cite{kipp09,gune07,jonathan10} and molecular electronics\cite{nitz03a,nitzan06}.

The Marcus theory relies in an essential way on electron-vibration interaction. The initial and final states of the electron transfer process are fully equilibrated polarons localized on different sites, and transitions between them is evaluated within the assumptions of transition state theory. Motion in an extended system is assumed to be a succession of hopping steps, each described as a Marcus process. In the other extreme limit, electronic motion in a frozen lattice, the electron moves within its energy band, most simply described using a tight binding model. In between these limits, electron-phonon interaction and band motion can change the electron's character from being weakly perturbed by electron-phonon scattering to polaronic motion whose discrete representation is the succession of hopping processes described above.

In the present paper we are interested in situations where electronic band motion competes on the same timescale with polaron formation, so that the dynamics of the latter process has to be considered explicitly. Such considerations are relevant to recently studied models of photovoltaic cells\cite{einax11,li12b}, where electrons (or holes) are injected at some location in the system and a useful process is defined by their absorption at another (e.g. an electrode surface). The yield of such processes, determined by the competition between electronic motion and loss processes\cite{li12b} (e.g. carrier recombination) is expected to be sensitive to electron-phonon interactions, and in particular to transient polaron formation.

Exact treatment of such coupled many-body systems is difficult, and it is tempting to resort to approximations such as the semiclassical mean field (Ehrenfest) dynamics. In this approximation the electronic wavefunction $\Psi (r,t)$ (r represents the electronic coordinates and t is the time) evolves under a time-dependent Hamiltonian defined by a classical nuclear trajectory, schematically represented by a nuclear coordinate $R(t)$, while the latter is obtained by solving the Newton equation for the nuclear motion with a potential in which the vibronic coupling $V(r,R)$ is replaced by its instantaneous expectation value $V(R,t)=\langle\Psi (r,t)|V(r,R)|\Psi (r,t)\rangle$. Such an approximation, essentially a dynamical extension of the Born-Oppenheimer (BO) approximation, is expected to perform well when the electronic motion is fast, throughout the relevant electronic subspace, relative to the nuclear dynamics. Its failure in describing processes in which transitions between BO electronic adiabatic states take place on the same timescale as nuclear motions is also well known. Indeed, in the analogous case of electron solvation in polar liquids such non-adiabatic processes have been addressed with the necessary accounting for the quantum nature of the nuclear motion\cite{szym05,turi05,prez97}, usually within the surface hopping methodology\cite{tully90,turi05,prez97}. Still, because Ehrenfest dynamics is so easy to implement and to use, it is of interest to assess its performance as an approximation to exact results in the context described above\cite{bonca95}. This is the purpose of the present paper. Using a model that is simple enough to solve up to any desired level of accuracy, we focus on two observables: the extent of the polaronic localization and its formation time, and compare results obtained from the semiclassical Ehrenfest dynamics approach to the exact, fully quantum, results. For model parameters that support polaron formation we find that, while the Ehrenfest calculation yields a similar final state as the exact one, it predicts a polaron formation time that is an order of magnitude longer than the exact result. This implies that Ehrenfest dynamics cannot be used as a reliable tool for assessing polaronic effects in such systems. This does not exclude its possible applicability in larger systems with higher temperatures with many more nuclear degrees of freedom, but indicates that its use should be exercised with caution and after performing suitable benchmark calculations.

We start by formulating the basic Hamiltonian model and comparing the different predictions of the quantum and the semiclassical descriptions in Section II.  In section III  we define the population operator and the population formation time. Numerical calculation and discussions are given in sections IV and V, and a conclusion follows.

\section{Theoretical Model}
We consider an n+1-site tight-binding free electron model (below we take n=4) coupled to a system of Harmonic oscillators. We assume that only one oscillator (henceforth referred to as the \textquotedblleft primary\textquotedblright vibration) directly couple to the electronic systems. The others (\textquotedblleft secondary\textquotedblright phonons) constitute a thermal bath that affects relaxation in the primary system. The Hamiltonian of the whole system is 
\begin{align}
  \label{h}
  H&= H_{S} +H_{B} +H_{SB}~,  \\
 \label{hs}
  H_S&=\sum_{l=0}^{4}\varepsilon_lc_l^\dagger c_l+V\sum_{l=0}^{3}(c_{l}^\dagger c_{l+1}+c_{l+1}^\dagger c_{l})
  +\hbar \omega_0 d_0^\dagger d_0+\alpha_2 c_{2}^\dagger c_{2} (d_0^\dagger+d_0)~,  \\
   \label{hb}
  H_B&=\sum_{s=1}^{\infty}\hbar \omega_s d^\dagger_s d_s~,  \\
  \label{hsb}
  H_{SB}&=\sum_{s=1}^{\infty} \lambda_s (d_0^\dagger d_s + d_{s}^\dagger d_{0})~. 
\end{align}

Here $H_S$ corresponds to the electronic system (described by the creation and annihilation operators for each site $l$, $c_l^\dagger$, $c_l$, together with the primary vibration, of frequency $\omega_0$ described by the creation and annihilation operators $d_0^\dagger$, $d_0$. $H_B$ describes the secondary phonon bath ($d_s^\dagger$, $d_s$ are the creation and annihilation operators for phonon of frequency $\omega_s$) and $H_{SB}$ is the coupling between the primary vibration and secondary phonons. $\varepsilon_l$ is the on-site energy level of site $l$, $V$ is the coupling parameter associated with electron tunneling between nearest neighbor sites. The parameters $\alpha_2$ and $\lambda_s$ correspond to the coupling between the electronic state at site 2 and the primary vibration, and between the primary vibration and secondary phonons, respectively.

\begin{figure}
\centerline{\includegraphics[width=16cm,clip]{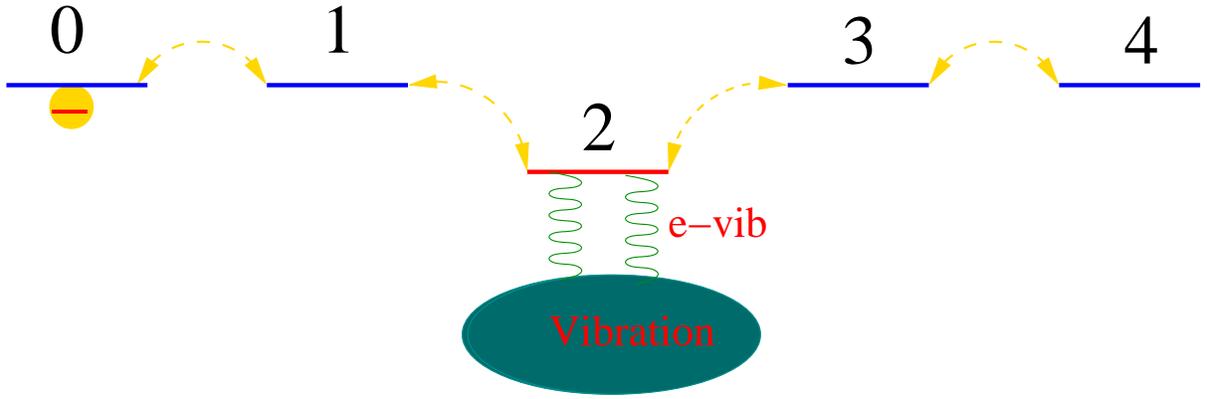}}
\caption{The model chain including 5 sites. The electron-vibration interaction occurs on the middle site 2.} 
\label{f.10}
\end{figure}

\subsection{The quantum approach}
In the quantum approach, the evolution described by the Hamiltonian (2) is treated essentially exactly\cite{bonca95}, using the basis set $\{|n,\nu>\}$ where n and $\nu$ denote the electronic state localized on site $n$ and the vibrational state $\nu$ of the primary oscillator. In the numerical calculation we truncate the set $\{\nu\}$ at some value, $\nu_{max}$ and test for convergence as $\nu_{max}$ increases.

The coupling to the thermal bath is treated in the master equation approach: The density matrix $\rho_T$ of the whole system is assumed to keep the form $\rho_T=\rho_S\otimes \rho_B$, where $\rho_{B}$, the density matrix of the thermal bath (secondary phonons), is assumed to remain in equilibrium at the ambient temperature. Then the quantum master equation (QME) for $\rho_S$ is
\begin{align}
\label{equ:master2}
i\hbar \frac{\partial\rho_S(t)}{\partial t}=
  [H_S, \rho_S(t)]-i\hbar\gamma_0[d_0^\dagger d_0 \rho_S(t)+\rho_S(t)d_0^\dagger d_0-2 d_0\rho_S(t)d_0^\dagger]/2~, 
\end{align}
where $\gamma_0$ is the vibrational relaxation rate induced by the vibration-phonon bath coupling. It is equal to the imaginary part of the vibration self energy $\Sigma$ given by
\begin{align}
\label{gamma}
\gamma_0(\omega)/2=\dfrac{1}{\hbar}\rm{Im} \{\Sigma_{vibration}(\omega) \}=\dfrac{1}{\hbar}\sum_s |\lambda_s|^2\delta(\hbar \omega-\hbar\omega_s)~~.
\end{align}

In the basis chosen, Eq.~\ref{equ:master2} takes the form
\begin{align}
\label{equ:master3}
i\hbar \frac{\partial\rho_{nv,n'v'}(t)}{\partial t}=
  [H_S, \rho_S(t)]_{nv,n'v'}-i\hbar\gamma_0(v+v')\rho_{nv,n'v'}\delta_{n,n'}/2+i\hbar\gamma_0 \sqrt{v+1}\sqrt{v'+1}\rho_{n,v+1,n',v'+1}~, 
\end{align}
where the first term in the right side of Eq.~\ref{equ:master3} comes from the contribution of primary system Hamiltonian $H_S$, and second part comes from the first two terms of the bracket on the right side of Eq.~\ref{equ:master2}, and the last term involves energy transfer from the higher to the lower vibrational levels\footnote{Because of the zero temperature and Bose-Einstein distribution function $N_B(\omega)=\dfrac{1}{e^{\hbar \omega/KT}-1}\equiv0$ (except for the ground vibrational level), the energy transfer only  happens from the higher levels to its nearest lower level. For non-zero temperature, there will be transfer from lower to higher levels.}. This equation will be solved numerically.

\subsection{The semi-Classical approximation}

 The dimensionless displacement of the single primary oscillator is  approximated in the semiclassical approximation by a time-dependent configuration $q(t)=<d_0^\dagger(t)+d_0(t)>$ as\cite{galp05b,li12c}
\begin{align}
\label{qt-grn0}
 q(t)=\dfrac{1}{\hbar}\int_{0}^{t}d\tau D^r(t-\tau) \alpha_2 <c_2^\dagger(\tau) c_2(\tau)>~,
\end{align}
arising from the interaction among the electron, the active vibration and the phonon bath; here $D^r$ is the retarded green function of the active vibration.

Equation 8 assumes that the displacement coordinate $q$ responds to the average electron population (in the present calculation, on site 2). This is a mean field description akin to the Ehrenfest approximation, that is to be tested in the calculations described below. 

Using the wide-band approximation
\begin{align}
D^r(\omega)=\dfrac{1}{\omega+\omega_0+i\gamma_0/2} -\dfrac{1}{\omega-\omega_0+i\gamma_0/2} 
\end{align}
and its Fourier transform
 \begin{align}
 \label{drt}
 D^r(t)=i[e^{(i\omega_0-\gamma_0/2)t}-e^{(-i\omega_0-\gamma_0/2)t}]=-2sin(\omega_0t) e^{-\gamma_0t/2}~,
 \end{align}
and substituting Eq.~\ref{drt} into Eq.~\ref{qt-grn0}, we get
\begin{align}
 \label{qt-grn}
  q(t)=-\dfrac{2}{\hbar}\int_{0}^{t}d\tau sin[\omega_0(t-\tau)] e^{-\gamma_0(t-\tau)/2} \alpha_2< c_2^\dagger(\tau) c_2(\tau)>~.
 \end{align}

Here $\gamma_0$ has been defined in Eq.~\ref{gamma}, and  neglecting the real part of the vibration self energy gives the solution for $q(t)$.

Finally replacing $d_0^\dagger+d_0$ in the electron-vibration coupling Eq.~\ref{hs} by $q(t)$ (Eq.~\ref{qt-grn}), we get the effective semiclassical electronic system Hamiltonian as\cite{galp05b,lakhno02, lakhno05}
\begin{align}
\label{heff}
H_{eff}=\sum_{l=0}^{\rm 4}\varepsilon_l c_l^\dagger c_l  +V\sum_{l=0}^{\rm 3}(c_{l}^\dagger c_{l+1}+c_{l+1}^\dagger c_{l} )+ F(t)c_2^\dagger c_2~,
\end{align}
with\footnote{Note that the last term in Eq.~\ref{heff} is proportional to $<c_2^\dagger c_2>\bullet c_2^\dagger c_2$. This is an artifact of the semiclassical approach.}
\begin{align}
\label{ft}
F(t)=\alpha_2 q(t)=-\dfrac{2\alpha_2^2}{\hbar}\int_{0}^{t}d\tau sin[\omega_0(t-\tau)] e^{-\gamma_0(t-\tau)/2} < c_2^\dagger(\tau) c_2(\tau)>~.
\end{align}    

The system density matrix $\rho_S$, which in this semiclassical approximation is derived from the Hamiltonian $H_{eff}$ in Eq.~\ref{heff}, can be solved using the Liouville equation
\begin{align}
\label{liou}
 i\hbar\dfrac{d\rho_S}{dt}=[H_{eff},\rho_S]~.
\end{align}

\section{population distribution, population formation time and electron-vibration coupling energy}

The on-site electronic population at any time t is 
\begin{align}
P_l(t)=< c_l^\dagger(t) c_l(t) >~, with~~ \sum_{l=0}^4 P_l(t)=1~. 
\end{align}
 
Since the time-dependent values $P_l$ oscillate, it is better to show these values using a coarse grained time-dependent average value
\begin{align}
 \label{average}
 \bar P_l(t)=\dfrac{1}{2\Delta T}\int_{t-\Delta T}^{t+\Delta T} d\tau P_l(\tau)~.
\end{align}
Below we use $\Delta T=50 fs$. In the following we will use ${\bar P}_l$ as the time-dependent average values for the populations.

The population formation time of $P_2$ can be defined as the time point at which population $P_2$ reaches a certain value. Experimentally one can define the \textquotedblleft formation time\textquotedblright as the time at which the target population reaches $\sim$ ($1-e^{-1}\approx 0.76$) of its final value\cite{lewis06}. Thus we use the criterion 
\begin{align}
\label{lifetime}
\bar P_2(\tau_p)= P^\infty_2 (1-e^{-1})~. 
\end{align}
$P^\infty_2$ is the time-averaged value of $P_2$ in the long time limit and $\tau_p$ is  the population formation time.

The electron-vibration coupling $E_P$ are
\begin{align}
\label{energyep}
E_p=\begin{cases}
 \alpha_2 <c_{2}^\dagger c_{2} (d_0^\dagger+d_0)>, & \mbox{quantum method,}  \\ 
  F(t)<c_2^\dagger c_2> ,& \mbox{semiclassical method.}
 \end{cases} 
\end{align}

\section{Numerical calculation}
For the initial numerical simulation, we set $\varepsilon_l=0$ for $l=0,1,3,4$ with $\varepsilon_2=-0.2eV$($\varepsilon_2$ is lower than the other site energies), $V=0.1eV$, $v_{max}=9$,  $\alpha_2=0.0707eV$,  $\omega_0=0.1eV$, $\gamma_0=0.04eV$. We will vary those parameters to examine more cases. All the results described below use the initial condition $c_n^\dagger c_n=\delta_{n,0}$ for the electronic state, and $\nu=0$ (ground vibrational level) for the primary phonon. For our purpose-comparing the quantum calculation and the mean-field semiclassical approximation-it is sufficient to consider the zero temperature case. 
 
For the quantum method, we need to set $\nu_{max}$ large enough to assure convergence of the calculation. Because we are at zero temperature, the energy transfer only happens from the higher levels to its nearest lower level and a smaller $\nu_{max}=3$ is enough (in the supporting information Fig. S1 shows that when $\nu_{max}$ is larger than 2 the average population ${\bar P}_2$ is similar). For non-zero temperature, there will be transfer from lower to higher levels, and larger $\nu_{max}$ will be needed.

\section{ Population localization and polaron formation}

\begin{figure}
\centerline{\includegraphics[width=0.9\textwidth,clip,angle=0]{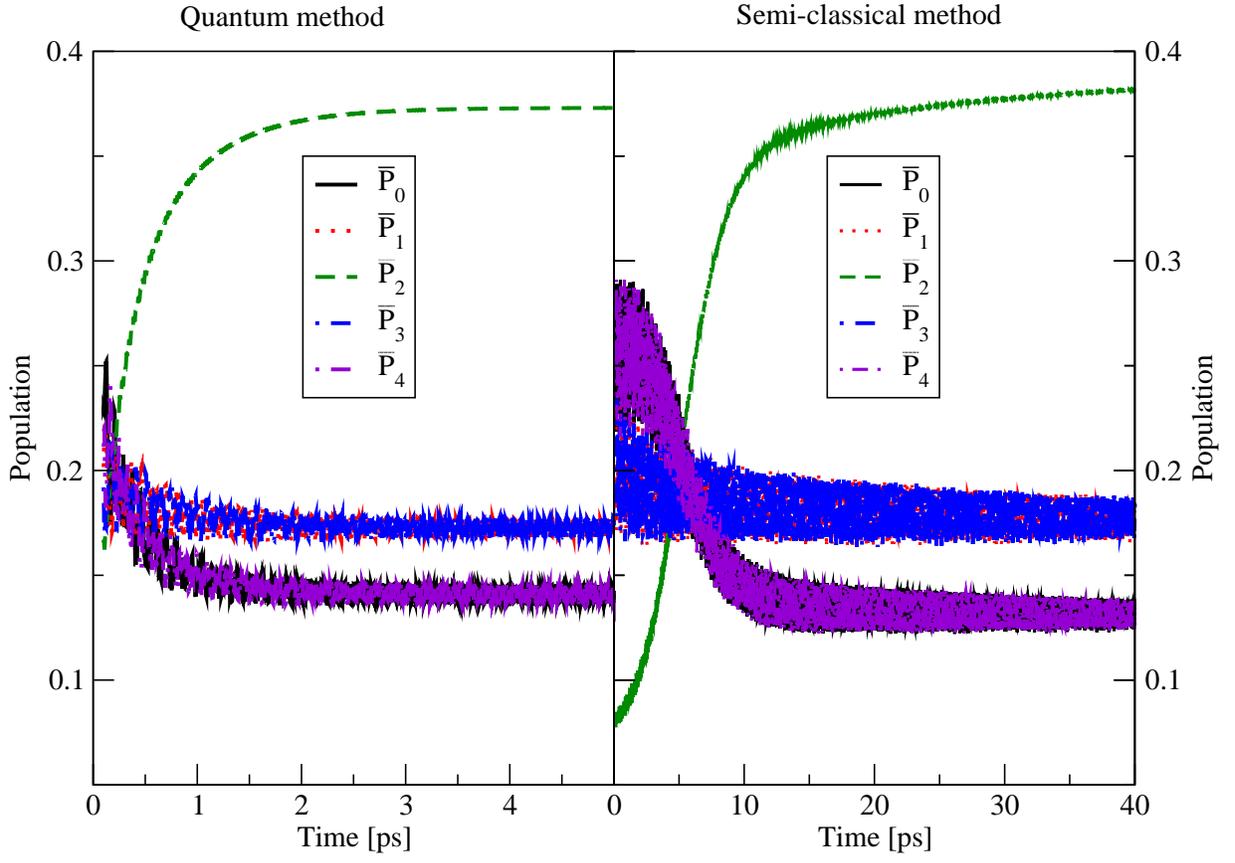}}
 \caption{Average population distribution on different sites $l$ ($l=0,1,2,3,4$) shown as a function of time.
$\varepsilon_l=0$ ($l=0,1,3,4$), $\varepsilon_2=-0.2eV$, $V=0.1eV$, $\alpha_l=0$ ($l=0,1,3,4$), $\alpha_2=0.0707eV$,  $\hbar\omega_0=0.1eV$, $\hbar\gamma_0=0.04eV$. Due to symmetry $\bar{P}_0$=$\bar{P}_4$; $\bar{P}_1$=$\bar{P}_3$.} 
 \label{f.2}
 \end{figure}

The electron-vibration coupling is allowed only at site 2 and the system is coupled to the bath by a single primary vibration. As shown in Fig.~\ref{f.2}, for a small $V$, much of the electron population will localize on site 2, forming a local polaron with larger population. The quantum results (Section II) are shown on left panel and the semiclassical results (section III) are displayed in the right panel. The respective results qualitatively similar:${\bar P}_2$ (polaron population) rises and saturates at values that are similar in both approaches. However the rise time obtained from the quantum calculation is much shorter than its semiclassical counterpart: ${\bar P}_2$ reaches its maximum value at 2 ps (picosecond) in the left panel, while in the right panel it takes about 15 ps. This difference demonstrates the shortcoming of the semiclassical approximation, or rather-its reliance on the mean field approximation: The localizing phonon moves on a potential surface that is substantially different from what it actually experiences once the localization process has started.

\subsection{Influence of the tunneling amplitude $V$}
The coupling $V$ between nearest neighbor sites is important for population localization. For small $V$ the coherent transfer between the different sites is weak and the population tends to localize, forming a polaron on site 2 which is coupled to the primary vibration and has been given a lower energy. For large $V$ the population tends to equalize on neighboring sites, showing an oscillatory behavior (here only  the average is shown ). As shown in Fig.~\ref{f.3}, the average population ${\bar P}_2$ on the impurity site decreases with increasing $V$, although the polaron is in principle formed on site 2. For $V=1.0eV$, the banding or delocalisation energy dominates over the electron phonon and impurity energies. In this limit, the charge population will tend  to reach a uniform average value. Recall that the original small-polaron model of Holstein\cite{holstein59,emin73,firsov75,sewell63} was developed for narrow-band (small $V$) materials.

\subsection{Effect of the electron-vibration coupling}
\begin{figure}
\centerline{\includegraphics[width=0.9\textwidth,clip,angle=0]{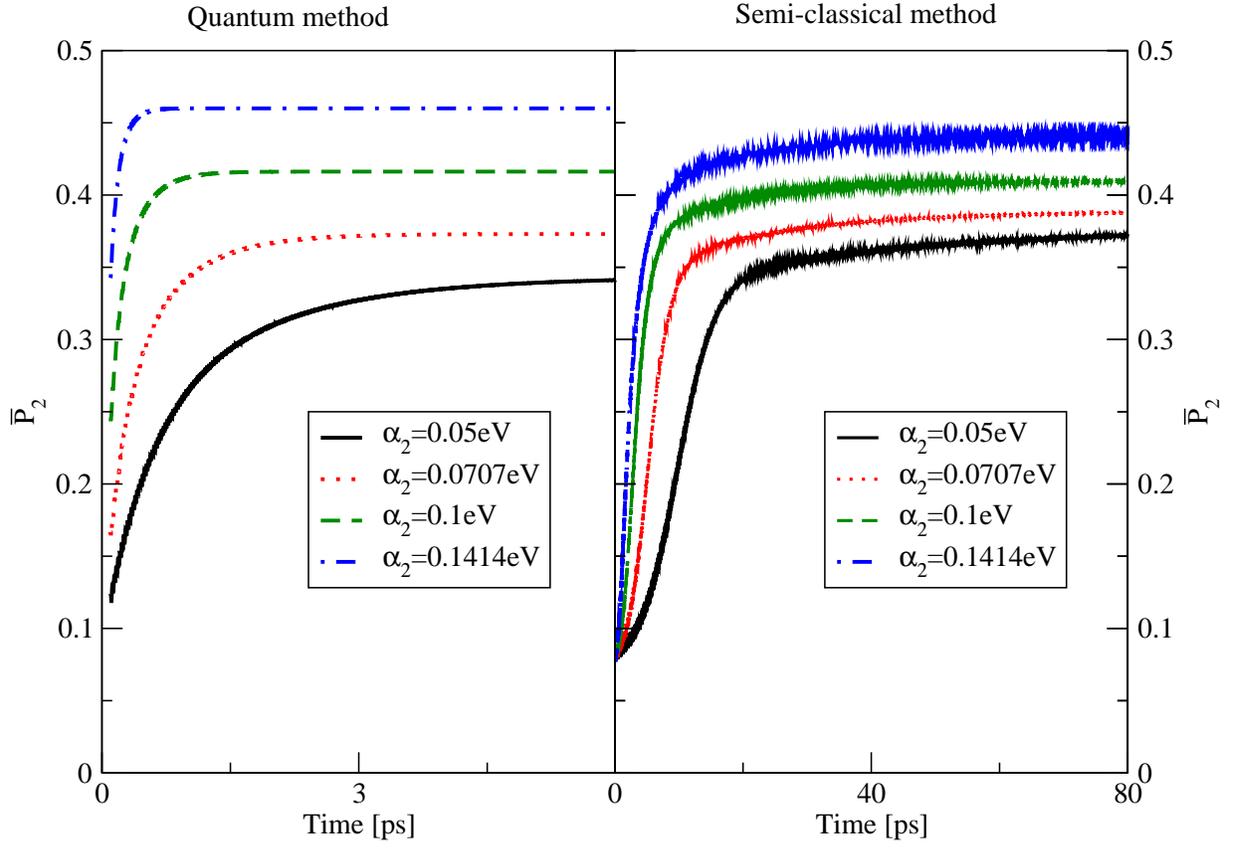}}
 \caption{Average population $\bar{P}_2$ shown as a function of time with different electron-vibration coupling parameter $\alpha_2$.
$\varepsilon_l=0$ ($l=0,1,3,4$), $\varepsilon_2=-0.2eV$, $V=0.1eV$, $\alpha_l=0$ ($l=0,1,3,4$),  $\hbar\omega_0=0.1eV$, $\hbar\gamma_0=0.04eV$.} 
 \label{f.4}
\end{figure}

$\alpha_2$ is the coupling strength between the electronic motion and the vibration at site 2. In the quantum method, electrons can evolve from the other sites to the vibronic states through this coupling as shown in the last term on the right side of Eq.~\ref{hs}.  With larger $\alpha_2$, more population will  transfer and form a polaron with large population on the middle site. For the semiclassical case, the same thing will occur  through Eq.~\ref{ft}, and some population will be localized to form a polaron. As shown in Fig.~\ref{f.4}, the population ${\bar P}_2$ in the steady state increases with this coupling strength. The more rapid population relaxation obtained in the quantum treatment occurs because in the quantum analysis, the actual population at site 2 is used, while in the semiclassical treatment, it is necessarily only the average population which drives the coupling.

\begin{figure}
\subfigure[quantum method]{\label{f.5-a}\includegraphics[width=0.85\textwidth,clip,angle=0]{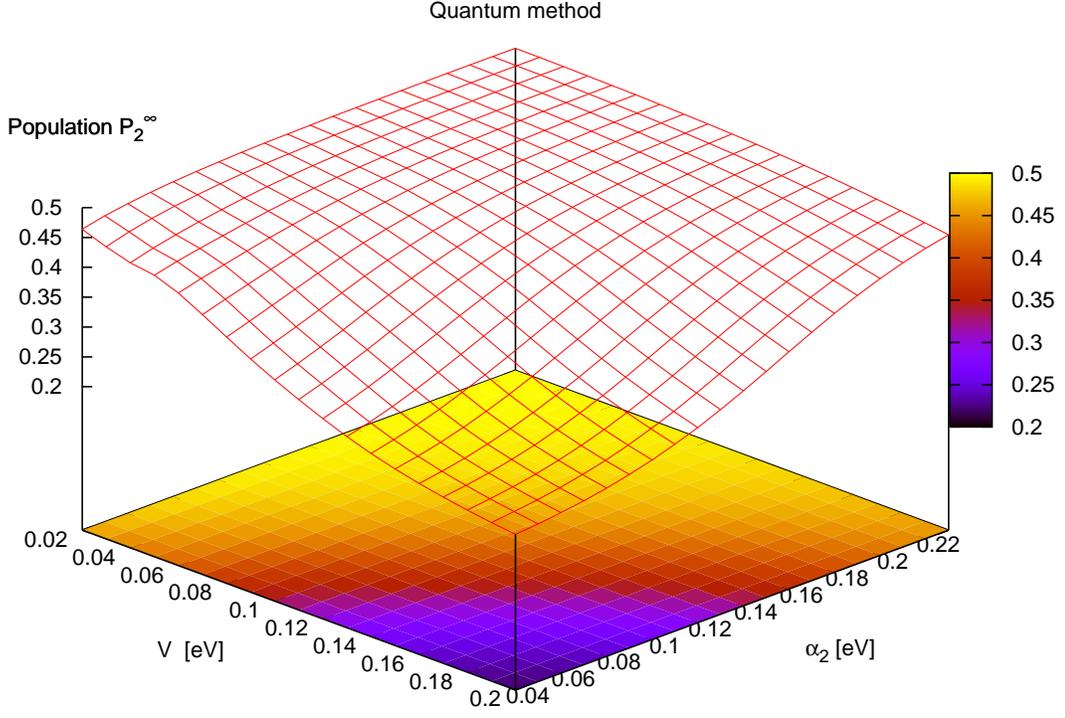}}
\subfigure[semiclassical method]{\label{f.5-b}\includegraphics[width=0.85\textwidth,clip,angle=0]{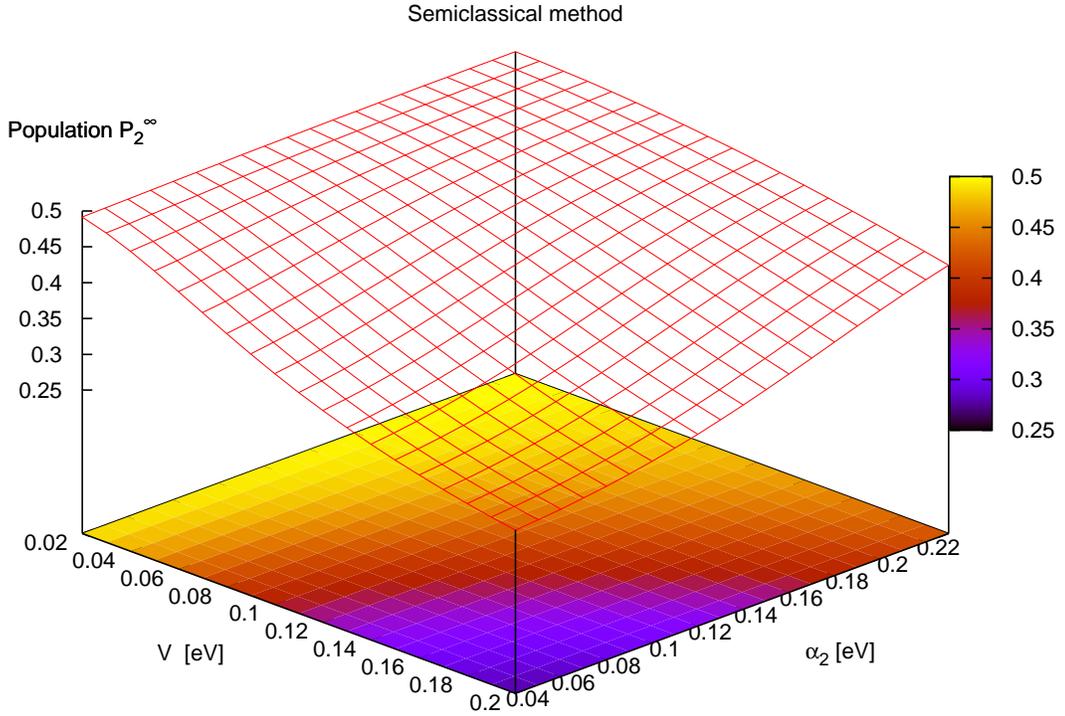}}
\caption{$P_2^{\infty}$ (population on site 2 in the steady state) shown as a function of the nearest neighbor site coupling parameter V and electron-phonon coupling $\alpha_2$. $\varepsilon_l=0$ ($l=0,1,3,4$), $\varepsilon_2=-0.2eV$, $\hbar \omega_0=0.1eV$, $\hbar\gamma_0=0.04eV$.  Quantum method  used for panel (a) and semiclassical method used for (b).}
\label{f.5}
\end{figure}

In Figs.~\ref{f.5-a} and ~\ref{f.5-b}, $P^\infty_2$ (defined in Eq.~\ref{lifetime}) is plotted in 3-D by changing the coupling $V$ and the electron-phonon coupling $\alpha_2$.  With a small $V$, $P^\infty_2$ is almost around 0.5\footnote{Note that localization with more than 0.5 the population on site 2 can indeed be achieved by allowing each site to be  coupled to its own  vibronic oscillator, even if these couplings constant are much weaker compared to site 2. This analysis, to be developed further in future,  leads to the tentative conclusion that the  \textquotedblleft true equilibrium distribution\textquotedblright, within the model itself, may not always be reached with oversimplified models. This question, namely under what circumstances can the system reach the  true ground  state (or excited state at finite temperature T) is an interesting one, and needs a lot more work to be done on it.}. Choosing larger values of $V$, decreases the site 2 population as can be expected, as shown in the corners of Figs.~\ref{f.5-a} and ~\ref{f.5-b}.  However increasing $\alpha_2$, the population builds up again, and we can say that the polaron population value is mainly determined by the parameters $V$ and $\alpha_2$.

\subsection{Comparison of  population formation times}
\begin{figure}
\centerline{\includegraphics[width=0.9\textwidth,clip,angle=0]{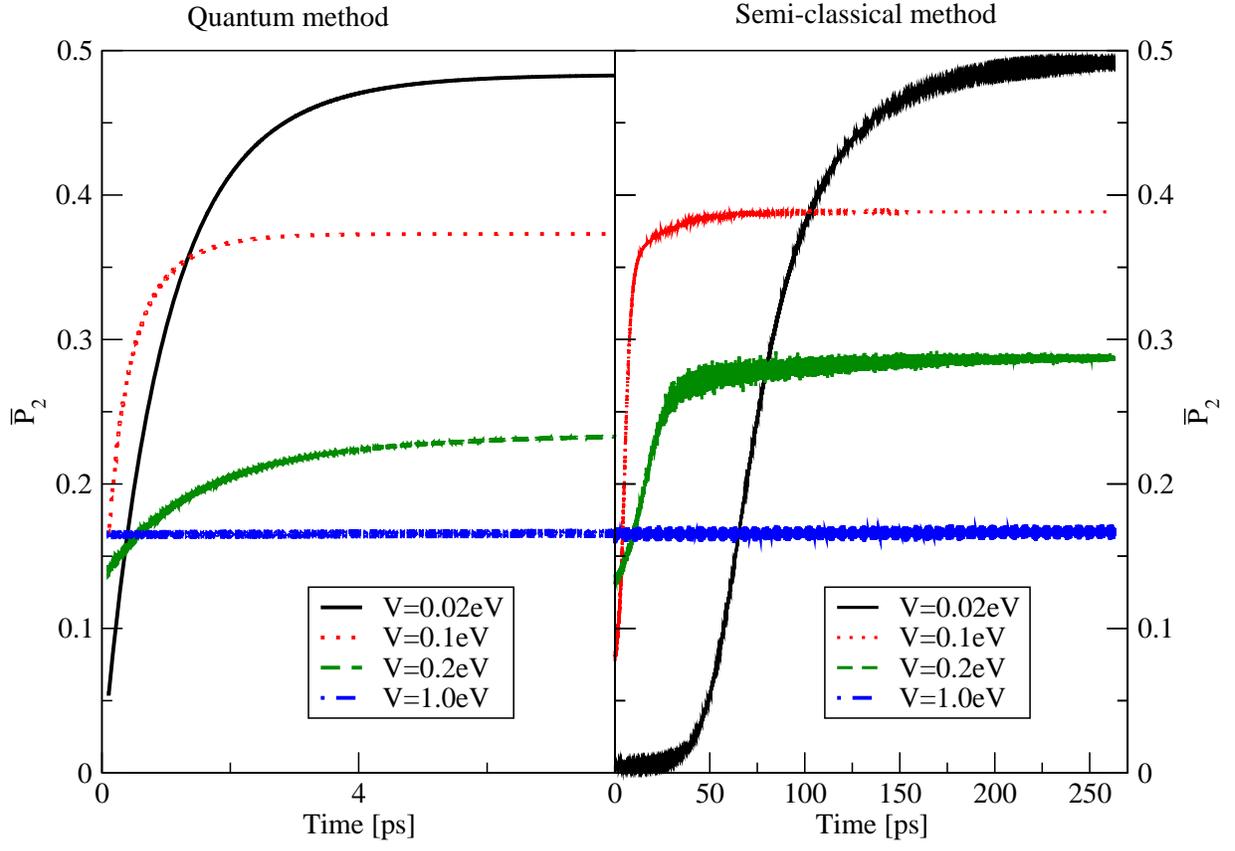}}
 \caption{Average population $\bar{P}_2$ as a function of time with different nearest neighbor tunneling parameter $V$.
$\varepsilon_l=0$ ($l=0,1,3,4$), $\varepsilon_2=-0.2eV$, $\alpha_l=0$ ($l=0,1,3,4$), $\alpha_2=0.0707eV$,  $\hbar\omega_0=0.1eV$, $\hbar\gamma_0=0.04eV$.} 
 \label{f.3}
 \end{figure}

\begin{figure}
\subfigure[quantum method]{\label{f.6-a}\includegraphics[width=0.85\textwidth,clip,angle=0]{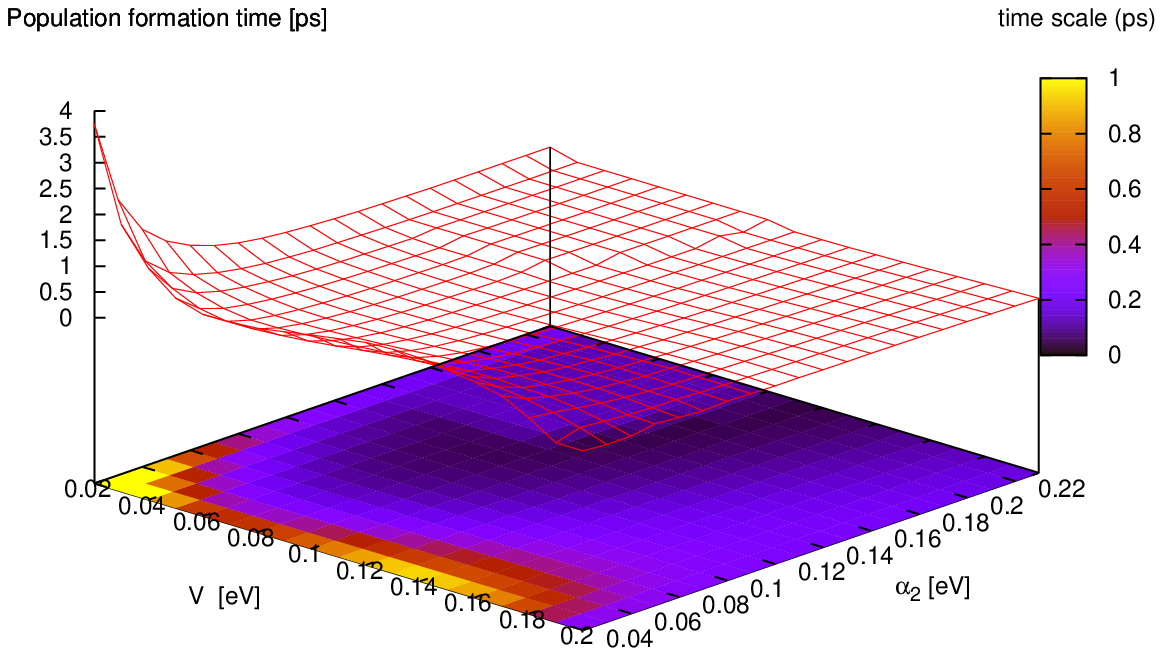}}
\subfigure[semiclassical method]{\label{f.6-b}\includegraphics[width=0.85\textwidth,clip,angle=0]{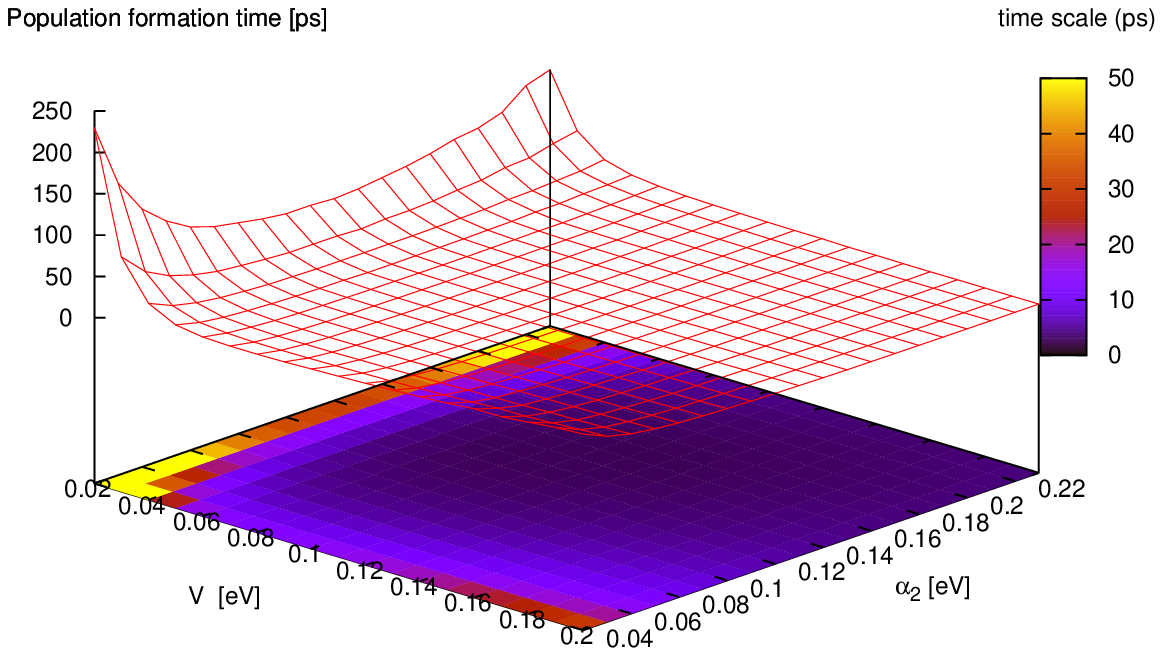}}
\caption{Population formatiom time for $P_2$ shown as a function of the nearest neighbor site coupling parameter V and electron-phonon coupling $\alpha_2$. $\varepsilon_l=0$ ($l=0,1,3,4$), $\varepsilon_2=-0.2eV$, $\hbar \omega_0=0.1eV$, $\hbar\gamma_0=0.04eV$. Quantum method  used for panel (a) and semiclassical method used for (b).}
\label{f.6}
\end{figure}

In Figs.~\ref{f.6-a} and ~\ref{f.6-b} the population formation time is shown in 3-D by changing both $V$ and $\alpha_2$ . The formation time obtained by the semiclassical method is longer than the quantum method.  The population formation time decreases with electron-phonon coupling $\alpha_2$ but then saturates. It also decreases with coupling $V$, but there is  a turnover in the  formation time beyond a V$\sim$0.1eV. This is due to the fact that for large $V$, the excitation relaxes into a delocalized population which is no longer strictly speaking a localized polaron. The population distribution is roughly constant in this limit as shown in Fig.~\ref{f.3}. The formation time should now be referred to simply as population relaxation time.

\subsection{The  short time dynamic of polaron formation  }
\begin{figure}
\subfigure[quantum method]{\label{f.7-a}\includegraphics[width=0.85\textwidth,clip,angle=0]{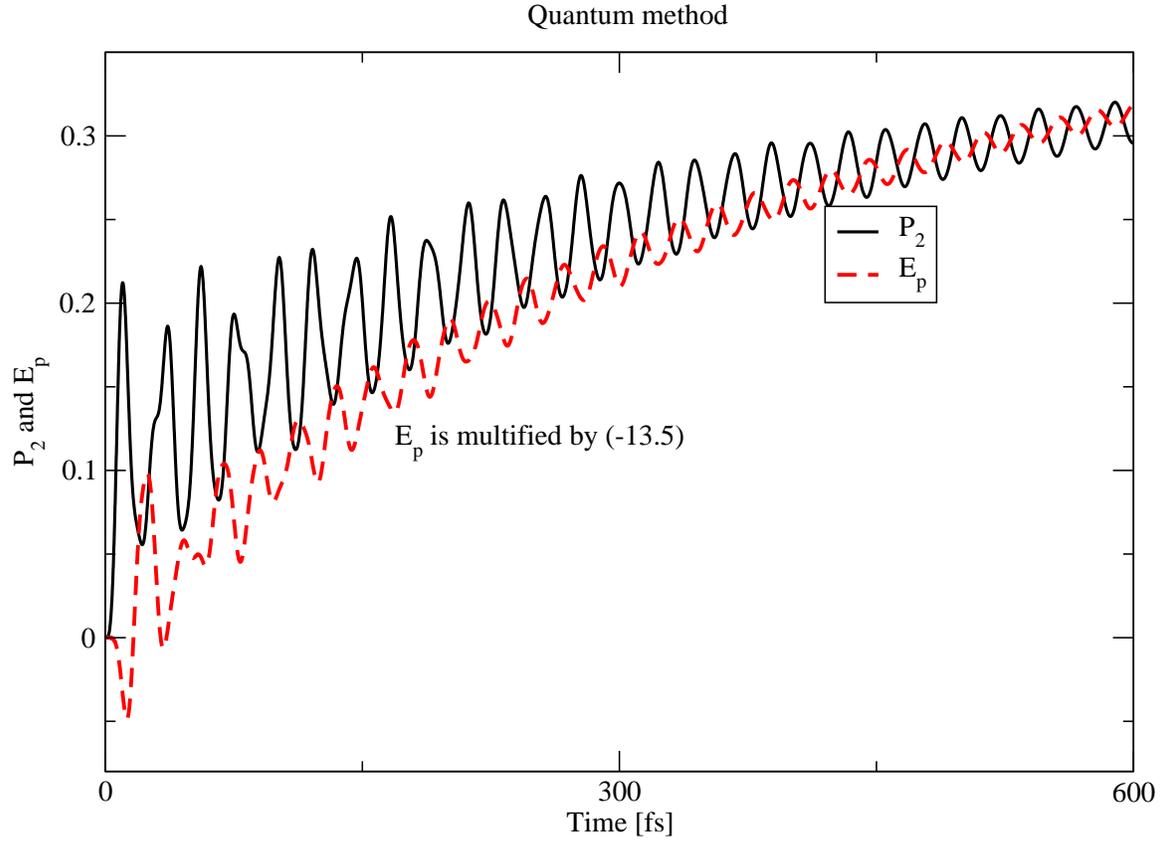}}
\subfigure[semiclassical method]{\label{f.7-b}\includegraphics[width=0.85\textwidth,clip,angle=0]{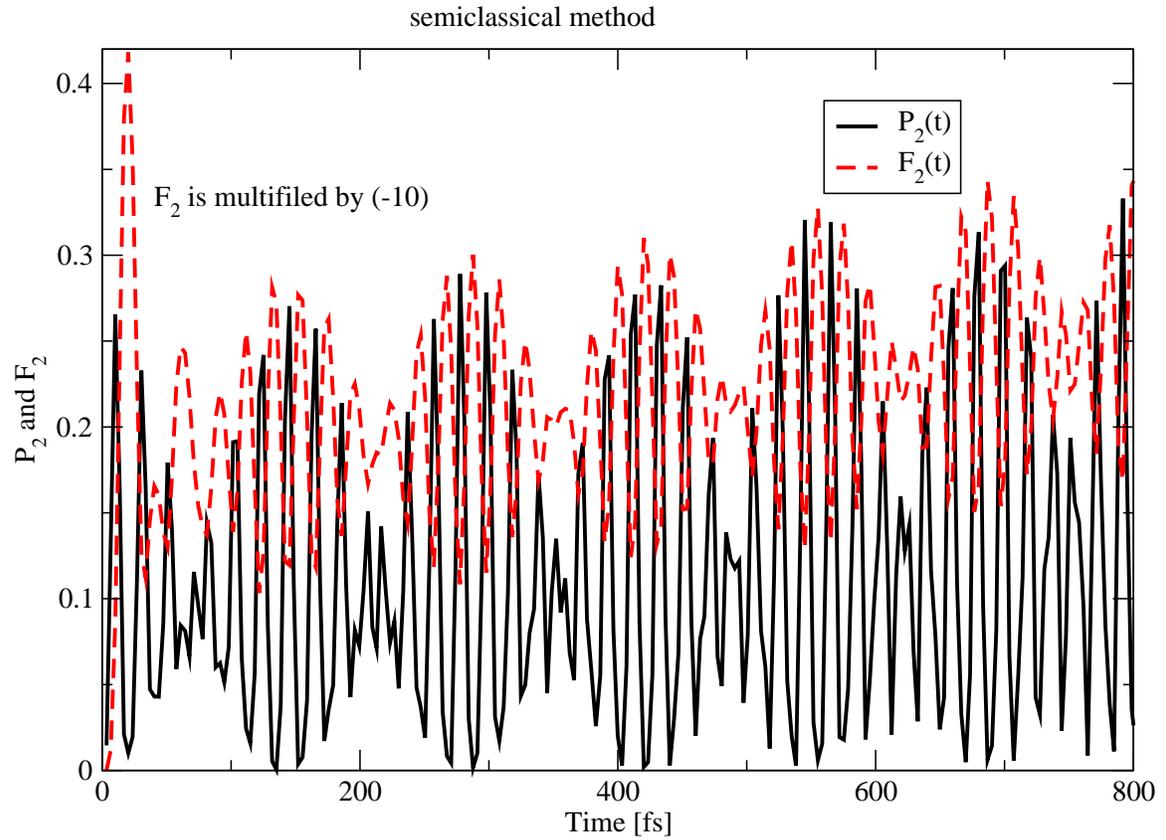}}
\caption{(a): Population $P_2$ and electron-vibration coupling energy $E_P$  are showing as a function of time;  (b): Population $P_2$ and the electron-phonon coupling energy $F_2(t)$ are showing as a function of time.
$\varepsilon_l=0$ ($l=0,1,3,4$), $\varepsilon_2=-0.2eV$, $\alpha_l=0$ ($l=0,1,3,4$), $\alpha_2=0.0707eV$  $\hbar\omega_0=0.1eV$, $\hbar\gamma_0=0.04eV$.  Quantum method  used for panel (a) and semiclassical method used for (b).}
\label{f.7}
\end{figure}

In Fig.~\ref{f.7-a} we compare the time-dependent dynamic processes for $P_2$ and $E_P$ (Eq.~\ref{energyep}). Both of them reach their steady state very quickly. When $P_2$ reaches its maximum, $E_p$ is at minimum, and vice versa.  This represents  the damping of a coherent oscillator where displacement and population keep their phase  delay throughout. The picture (Fig.~\ref{f.7-b}) is somewhat different in the semiclassical case, $P_2$ and its $F_2$  are also  delayed but we now see beats due to the interference of waves scattering from an oscillating potential.  Both quantities  reach their steady  state  more  slowly in this case.

\section{Conclusion} 

In this paper we have compared a fully quantum and semiclassical model. The latter based  on Ehrenfest dynamics for the calculation of the polaron formation process. We used a 1-dimensional tight binding model that includes  electron-phonon coupling but only one site, the so called polaron trap, the \textquotedblleft trap phonon\textquotedblright is in turn coupled  to a  thermal bath which allows the system to relax into  \textquotedblleft an equilibrium\textquotedblright.

The results from both methods show qualitatively similar behavior, however with markedly different timescales:
the population formation time obtained from the quantum calculation can be 10 times faster compared to the semiclassical method. This discrepancy becomes smaller with increasing intersite coupling $V$ (for large $V$ no localized polaron is formed) since  the classical limit is reached for $V\gg\omega_0$.

The different relaxation times obtained in the quantum and the semiclassical calculations result from the use of mean field approximation in the latter. In this approximation, the primary oscillator responds to the average occupation of site 2 which effectively makes it move on a potential surface that is markedly different (less binding) than the one it experiences once localization is initiated. Localization on this average potential is slower. This averaging assumption may be justified when the electron motion is much faster than the vibration, that is, for large $V$ ($V\gg \hbar \omega_0$). 
Another minor difference between the two calculations is the use of a master equation in the quantum calculation, and an essentially equivalent Langevin equation in the semiclassical one, to describe the relaxation of the primary phonon. These relaxation schemes are equivalent, provided that care is taken to use parameters that imply the same relaxation rate in both cases, as was done here.

The conclusion would also apply to the \textquotedblleft exact\textquotedblright simulations of Kopidakis et al\cite{kopida95} and the many other papers where the vibrations are also treated with semiclassical dynamics. It would seem that depending on electron bandwidth, the formation time is considerably underestimated in these works, perhaps by an order of magnitude or more. The semiclassical  results are essentially in agreement with the conclusions reached by Emin and Kriman\cite{emin86,emin12} using the Holstein diatomic polaron lattice model. These authors showed that population localization and polaron formation depend critically on the ratio of the tunneling energy $V$ and the width of the Bloch phonon dispersion, which effectively plays the role of the dissipation term since outward traveling phonon Bloch waves will not return. One final but very important point. We have shown that for simple  relaxation models, the true thermodynamic  equilibrium is not necessarily the one reached in the steady state. The state reached with arbitrary start conditions  can, in terms of energy, be a metastable state the populartion on site 2 (lower energy ) never exceeds however small we make V.  Introducing even a  small coupling on the other sites then allows the system to reach the true ground state and now the population can climb up to 1. This very interesting  point needs to be investigated in detail  and in particular at finite temperature.

\textbf{Acknowledgement}

This work was supported by the Non-Equilibrium Energy
Research Center (NERC) which is an Energy Frontier Research
Center funded by the U.S. Department of Energy, Office of
Science, Office of Basic Energy Sciences under Award Number
DE-SC0000989. MR thanks the chemistry division of the NSF (CHE-1058896) for support. The research of A. N. is supported by the Israel Science Foundation Grant No. 1646/08, the U.S.-Israel Binational Science Foundation, and the European Research Council under the European Union's Seventh Framework Program (FP7/2007-2013; ERC grant agreement $n^\circ$ 226628). The authors would like to thank Boris D. Fainberg for his insightful remarks.

 \newpage

\end{document}